# Direct Observation of Micro Structures on Superconducting Single Crystals of $K_xFe_{2-y}Se_2$

Masashi TANAKA[1,a], Hiroyuki TAKEYA[1], and Yoshihiko TAKANO[1,2]

[1]*MANA, National Institute for Materials Science, 1-2-1 Sengen, Tsukuba, Ibaraki 305-0047, Japan*

[2]*University of Tsukuba, 1-1-1 Tennodai, Tsukuba, Ibaraki 305-8577, Japan*

[a]Corresponding author: Masashi Tanaka,

E-mail: Tanaka.Masashi@nims.go.jp

Postal address: National Institute for Materials Science,

1-2-1 Sengen, Tsukuba, Ibaraki 305-0047, Japan

Tel.: (+81)29-851-3354 ext. 2976

**Abstract**

Potassium intercalated FeSe has been reported as a superconductor with superconducting transition temperatures ($T_c$'s) of 30-48 K. However, there is no clear answer to identify the relationship between the surface morphology, compositional ratio and its crystal structure. This report directly reveals the correspondence among these three characteristics in single crystals with a $T_c$ onset of around 44 K using a micro-sampling technique. Island-like parts on the surface of the crystals clearly have the $K_xFe_2Se_2$ structure with perfect FeSe layers, which is formed in conjunction with the $K_2Fe_4Se_5$ phase. This results in the appearance of the $T_c$ onset of 44 K.

After the discovery of superconductivity in potassium intercalated FeSe,[1] tremendous progress has been made in the study of related $A_x$Fe$_{2-y}$Se$_2$ ($A$ = K, Rb, Cs, Tl/Rb, Tl/K) systems which exhibit relatively high superconducting (SC) transition temperatures $T_c$ ~30 K.[2-7] Formed in isolation from the 30 K phase, a trace amount of a 44-48 K SC phase was also occurring in some samples[8,9] or under high pressure.[10,11] Alkali or alkaline-earth intercalation into FeSe via a liquid ammonia route has been reported to show superconductivity at $T_c$ ~44-46 K.[12,13] And a high $T_c$ is also observed in ultrathin FeSe films.[14,15] These reports imply that the superconductivity as high as 40-48 K is a potential property of the FeSe layer structure.

The K$_x$Fe$_{2-y}$Se$_2$ is one of the most investigated materials among the metal intercalated FeSe SC families. Its crystal structure is composed of a basic FeSe layer which incorporates some Fe deficiency mainly due to the conservation of the charge neutrality. Many studies have shown that intrinsic phase separations occur in the material, leading to the coexistence of a SC phase with the ThCr$_2$Si$_2$-type structure (122-phase) and an Fe-vacancy ordered insulating K$_2$Fe$_4$Se$_5$ phase with $\sqrt{5} \times \sqrt{5} \times 1$ superstructure (245-phase), as shown in Fig. 1(a)-(d).[16-30] *In-situ* observation with scanning electron and transmission electron microscopies at elevated temperatures have shed some light on a potential growth mechanism of the SC phase in single crystals.[31,32] Recently, we have discussed the mechanism underlying the production of the higher-$T_c$ (~44 K) phase in the K$_x$Fe$_{2-y}$Se$_2$ single crystals by means of *in situ* high-temperature single-crystal X-ray diffraction together with the evaluations of the surface morphology.[33] The X-ray diffraction pattern showed a reversible 245-122-phase transition with a transition temperature around 275°C. The superstructure spots in the *hk0* section corresponding to the 245-phase disappeared above 280 °C as shown in Figs. 1(e), (f). It was concluded that the higher-$T_c$ phase was generated by concentrating Fe onto the 122-phase with a driving force from a formation of the 245-phase.[33]

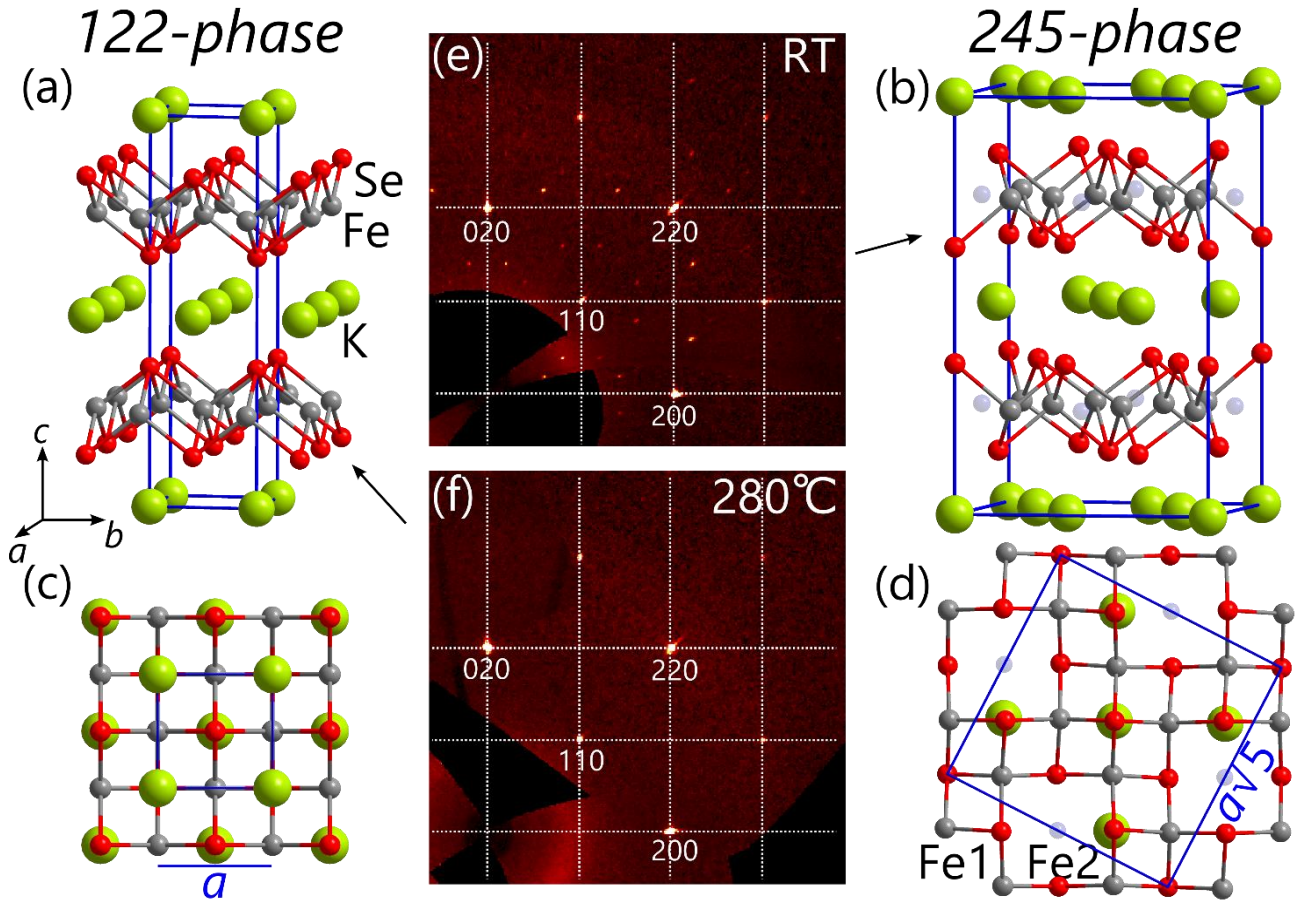

Figure 1. Schematic illustrations of unit cells for (a) the 122-phase, $K_xFe_2Se_2$, and (b) the Fe-vacancy ordered 245-phase, $K_2Fe_4Se_5$ ($K_{0.8}Fe_{1.6}Se_2$), structures. (c), (d) The corresponding FeSe lattice viewing along *c*-axis direction. Transparent atoms indicate the Fe-vacancy in the structure. Single crystal X-ray diffraction spots in the *hk0* section at room temperature (e), and 280 °C (f). All the Miller indices are given on the basis of the 122-phase structure cell with lattice parameters $a = b$ ~3.9 Å. There are some Bragg peaks of the superstructure in (e) arising from the 245-phase with lattice parameters $a = b$ ~8.7 Å.

However, the X-ray diffraction and scanning electron microstructural composition analyses always average the two phases on the measured single crystals. On the other hand, the transmission electron microscope (TEM) observation is highly localized and sometimes relies on the adventitious findings of the target sample. It might not be appropriately related to the micro-scale sections of the single crystals. It is necessary to show evidently the correspondence between the crystal structure, compositional ratio and surface morphology of the single crystals. In this study, we have clearly demonstrated the direct observations by combining TEM measurements with Hitachi micro-sampling technique. The appearance of superconductivity in K-Fe-Se system is discussed on the basis of the measurements and previous X-ray diffraction studies.

Target single crystals of K-Fe-Se system were prepared using a similar technique to the one-step-method.[34] Powders of $K_2Se$, Fe, and Se grains were mixed with a nominal composition of $K_{0.8}Fe_2Se_2$ in an Ar atmosphere. The starting mixtures were placed in an alumina crucible, and were sealed in evacuated quartz tubes. The quartz tube was heated to 900°C in 5 h, the temperature was then held for 12 h, followed by cooling to room temperature at a rate of 7°C/h. All the sample preparation was performed in an Ar-filled glove box. The temperature dependence of electrical resistivity was measured down to 2.0 K, with a Physical Property Measurement System (PPMS, Quantum Design) using a standard four-probe method with constant current mode. The electrodes were attached in the *ab*-plane with silver paste. The temperature dependence of magnetization was measured using a SQUID magnetometer (MPMS, Quantum Design) down to 2 K under a field of 10 Oe, and the field was applied parallel to the *c*-axis. Back scattered electron (BSE) and secondary electron (SE) images were observed using a scanning electron microscope (JEOL, JSM-6010LA). The sample fabrication for TEM observation was carried out by using focused ion beam (FIB) apparatus (HITACHI, NB5000) with Hitachi Micro-Sampling technique.[35] The electron diffraction patterns were observed using HITACHI HF-3300 with an acceleration voltage of 300 kV. The elemental distribution mapping of energy dispersive X-ray (EDX) analysis was carried out by scanning transmission electron microscope (STEM) with spherical aberration correction (HITACHI, HD-2700). The sample was avoided from air exposure during the sample exchange among each apparatus by using an air protection holder.

Figure 2(a) shows the temperature dependence of resistivity for the obtained single crystals. The crystal shows a SC transition with the onset temperature of $T_c$ at 44 K and zero resistivity at around 33 K. The transition was suppressed by applying a magnetic field. Fig. 2(b) is an enlargement scale of the temperature dependence of magnetic susceptibility ($\chi$) in the zero-field-cooling (ZFC) mode. It shows the Meissner signal at ~43 K and ~33 K compatible with the resistivity drop. The shielding volume fractions of ~43 K and ~33 K phases were roughly estimated at 2 K about ~1% and ~10%, respectively.

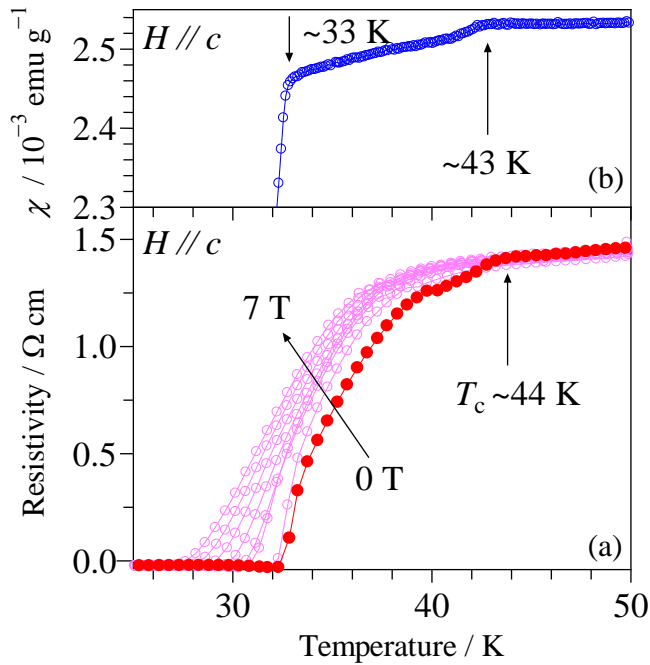

Figure 2. Temperature dependences of (a) resistivity and (b) magnetic susceptibility $\chi$ in enlarged scale of the obtained single crystals. The magnetic fields were applied parallel to the $c$-axis of the crystal.

The BSE and SE images of the same position of the obtained single crystals are shown in Fig. 3. It can be obviously seen that the crystals are composed of two different contrasted regions with clear boundaries. The bright island-like contrasted regions in the BSE image correspond to more iron and less potassium contents compare to those of the other dark regions (out of the island-like part).

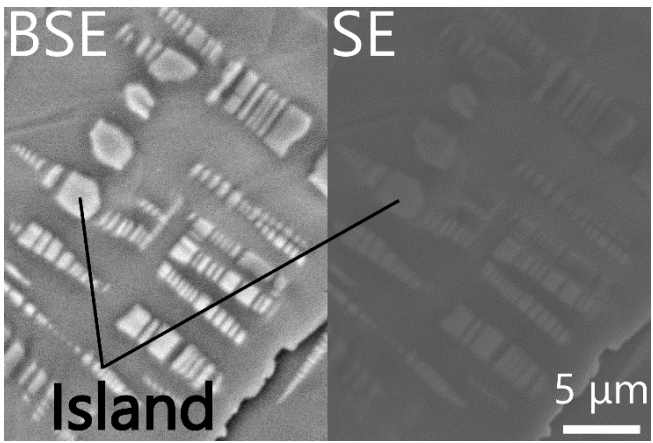

Figure 3. BSE and SE images of the obtained single crystals.

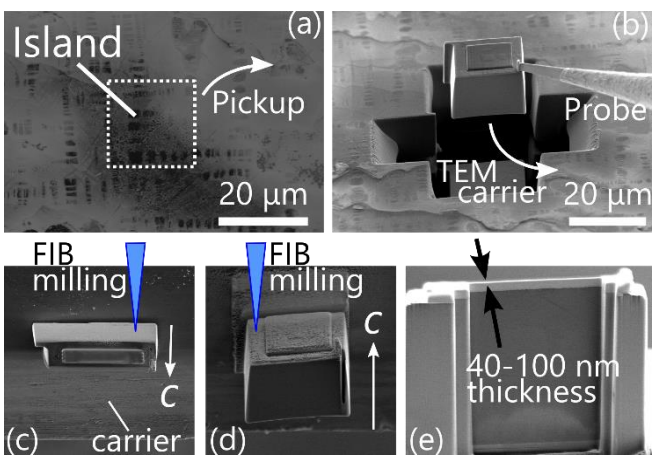

Figure 4. Procedures of the micro-sampling technique. Separately picked up specimens were attached to the TEM carrier with different orientations of the crystals like as (c) and (d). The micro-samples are thinned down to 40-100 nm in both orientations after attaching to the TEM carrier.

These facts indicate that there are two crystal structures, two SC transitions, and two compositional ratios on the same sample single crystals. It is necessary to identify the correspondence between the $T_c$, the surface morphology, compositional ratio and its crystal structure. To reveal this correspondence, an electron diffraction measurement and high resolution elemental mapping on a specific position of the sample crystals would become crucial probes. The specimen preparations are achieved by HITACHI micro-sampling technique using FIB, and the procedures are shown in Fig. 4. Figure 4(a) is a scanning ion microscope image of a surface on the as-obtained single crystals. Clear island-like morphology is observed in the image. A section including the island-like parts was firstly covered with a protective layer of carbon by FIB assisted deposition. Then a portion of the sample (micro-sample) was separated from the bulk sample and picked up by bonding a tungsten (W) micromanipulator probe with W deposition (Fig. 4(b)). These operations were carried out both in the cross sectional and *ab*-plane directions of the crystal structure. The micro-samples with different orientation of the crystals were mounted onto an edge of a TEM carrier by the W deposition, and the probe was cut off by FIB milling (Fig. 4(c), (d)). Finally, the micro-samples were thinned down be suitable for TEM observation as the thickness to be ~40-100 nm. The fabricated specimens were then transferred to STEM or TEM apparatuses using an air protection holder for observation of elemental mapping with high spatial resolution and electron diffraction patterns.

EDX elemental mapping images for the fabricated micro-sample in the *ab*- and cross-sectional planes of the obtained crystals are shown in Figs.5(a)-(c) and Figs.5(d)-(f), respectively. The distribution difference of K and Fe atoms are well resolved to island-like shape (Figs. 5(a), (b)), as which appears in the SEM/BSE images. The Fe-rich-K-less regions are arranged in stripe-like patterns in the cross-sectional plane, and they disappeared in around half of the viewings (Figs. 5(d), (e)), indicating three-dimensional network of the island-like parts, similarly observed in a stripe patterns of Ref.31. On the other hand, the Se atoms are homogeneously distributed in the whole of the observed area (Figs. 5(c), (f)). The highly convergent electron beams of STEM and the sample thickness effectively improve the spatial resolution of EDX analysis compare to that of the bulk sample. The compositional ratio of the island-like and the other parts are estimated to be $K_{0.43(5)}Fe_{2.02(6)}Se_2$ and $K_{0.79(4)}Fe_{1.77(6)}Se_2$ in the *ab*-plane, and $K_{0.48(3)}Fe_{1.91(6)}Se_2$ and $K_{0.63(7)}Fe_{1.72(4)}Se_2$ in the cross-sectional plane, respectively.

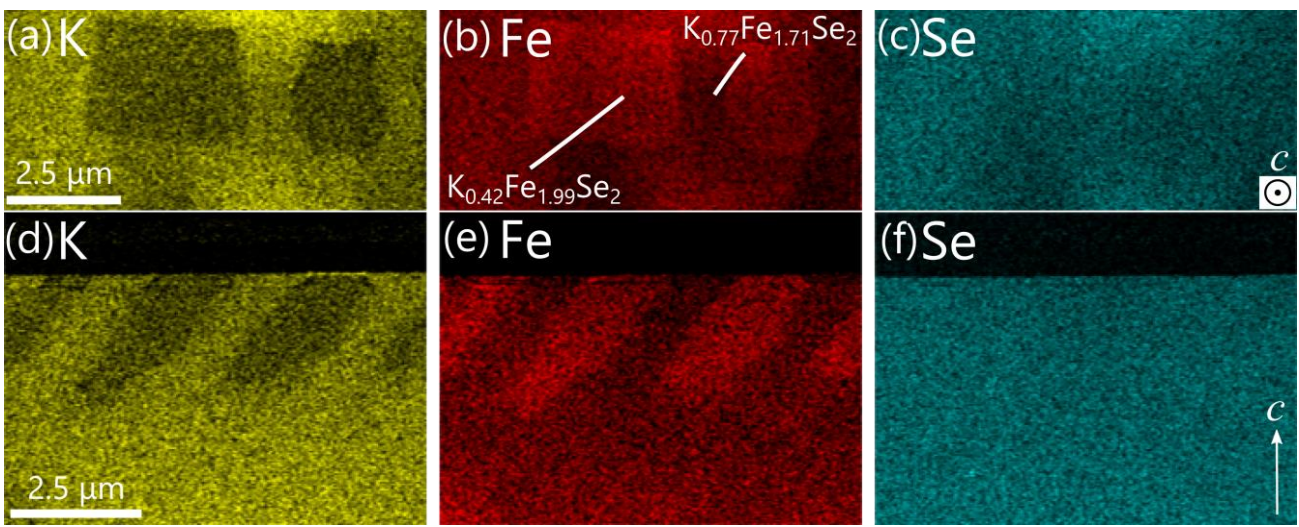

Figure 5. EDX elemental mapping in the *ab*-plane (a)-(c) and cross sectional plane (d)-(f) of the fabricated micro-samples.

The electron diffraction patterns of the fabricated micro-sample were measured with the electron-beam incidence normal to the *ab*-plane; these are shown in the Figs. 6(a), (b). The pattern indicated that the micro-sample basically has a 4-fold symmetry axis. All diffraction spots can be indexed by tetragonal unit cell, although the diffractions from out of the island-like parts always show superstructure spots around the Bragg spots of basic structure as shown in Fig. 6(b). This pattern is identical to the room temperature X-ray diffraction pattern as displayed in (Fig. 1(e)). Note that the diffractions from anywhere on the island-like parts did not show signatures of a superstructure spot. Namely the "islands" are composed solely of the 122-phase structure as observed in the higher temperature region above 275°C, on the other hand, the "out of the islands" have the 245-phase structure.

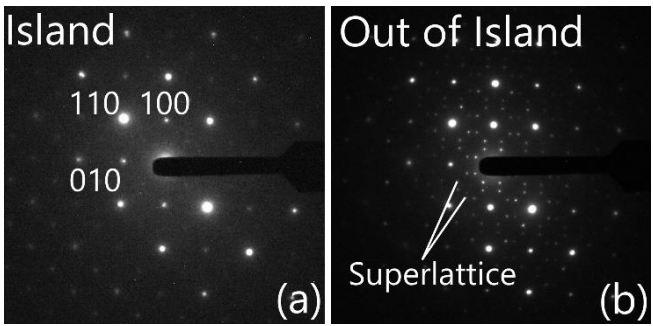

Figure 6. Electron diffraction patterns on (a) the island-like part, (b) out of the island-like part with the incident beam normal to the *ab*-plane of the crystals. The Miller indices in (a) are given on the basis of a tetragonal unit cell.

The crystals showed highly diffused patterns except for the *hk0* plane in the single crystal X-ray diffraction measurements. In the TEM measurements, stacking fault like features along its *c*-axis direction were also observed only on the island-like parts in the cross-sectional directions, as which appears in horizontal striped lines (data not shown). The highly diffused patterns in the X-ray diffraction measurements may be attributed to a break in the long-range order along the *c*-axis caused by an unregulated stacking of layers during the growth of the single crystal. The 122-phase accumulate the elastic strain along the *c*-axis during the crystal growth process reflecting the *c*-value difference with the 245-phase as is expected from the X-ray diffraction measurements.[33]

These facts give us an understanding of a direct correspondence between the surface morphology, compositional ratio and the crystal structures. The island-like parts of the single crystal has a crystal structure of the 122-phase with the compositional ratio of $K_{0.43(5)}Fe_{2.02(6)}Se_2$. Namely it contains "perfect" FeSe-layers in its crystal structure. On the other hand, the outer parts of the island correspond to the 245-phase with a slightly richer Fe content compared to the stoichiometry of $K_2Fe_4Se_5$. The excess Fe plays a role of carrier doping to the insulating 245-phase, suggesting that it may affect to conducting nature in the transport measurement, and which results in a finite resistivity at room temperature.

When the 245-phase is grown below 275°C, the excessively accommodated Fe has expelled from the structure to create the Fe-vacancy sites. The migrated excess Fe is concentrated onto small isolated island-like regions forming perfect FeSe-layers. It is reasonable that the 122-phase with perfect FeSe layers shows the higher $T_c$ onset of ~44 K in the electrical or magnetization measurements, taking into account a potential SC transition of FeSe up to ~48 K. However, there still remains a question related to the ~33 K superconducting phase. It should be clarified that which part corresponds to the ~33 K phase in the single crystals.

In summary, the relationship among surface morphology, compositional ratio and crystal structure of the K-Fe-Se superconducting system were directly identified by using micro-sampling technique. Island-like parts of the crystals clearly showed diffraction spots identical to those of the 122-phase structure. Out of the island-like parts have a crystal structure of the 245-phase with excess Fe compare to its stoichiometry. The appearance of higher $T_c$ onset of 44 K is attributed to the formation of 122-phase with perfect FeSe layers assisted by the growth of 245-phase. The micro-sample electrical measurement is also required by using the same sample pick up system to reveal the structural evidence surrounding the two $T_c$ onsets.

**Acknowledgment**

We would like to thank M. Shirai, K. Ito, H. Matsumoto and H. Yamada of Hitachi High-Technologies Corporation for their experimental supports. And we also show a gratitude to Dr. S.J. Denholme of Tokyo university of science for proofreading of the manuscript.